\documentclass[a4paper,oneside,twocolumn,notitlepage]{extarticle_ecoc}
\UseRawInputEncoding
\usepackage{bm}
\usepackage{amsmath}
\usepackage{graphicx}

\makeatletter

\pdfpageheight\paperheight
\pdfpagewidth\paperwidth

\usepackage{enumitem}

\usepackage{ecoc}

\makeatother

\usepackage[style=ieee]{biblatex}
\begin{document}
\title{Joint Subcarrier Phase Recovery for Nonlinearity Mitigation}
\author{Marco Secondini\textsuperscript{(1,2)}, Stella Civelli \textsuperscript{(3,1)}}

\maketitle
\begin{strip}
\begin{author_descr}

\textsuperscript{(1)} TeCIP Institute, Scuola Superiore Sant\textquoteright Anna,
56124 Pisa, Italy, marco.secondini@santannapisa.it.

\textsuperscript{(2)} PNTLab, Consorzio Nazionale Interuniversitario
per le Telecomunicazioni (CNIT), 56124 Pisa, Italy.

\textsuperscript{(3)} CNR-Istituto di Elettronica e di Ingegneria
dell\textquoteright Informazione e delle Telecomunicazioni (CNR-IEIIT),
56122 Pisa, Italy.
\end{author_descr}

\end{strip}

\begin{strip}
\begin{ecoc_abstract}
We propose a low-complexity phase recovery scheme that simultaneously
mitigates laser phase noise and fiber nonlinearity across several
subcarriers. In a long single-span link with Raman amplification,
the scheme achieves 0.9\,dB gain with 99 real multiplications per
complex symbol. \textcopyright 2026 The Author(s)
\end{ecoc_abstract}
\end{strip}

\section{Introduction}

Fiber Kerr nonlinearity limits the performance of modern optical fiber
communication systems. Its modelling and mitigation have been widely
investigated, primarily in the context of long-haul systems employing
coherent detection and advanced digital signal processing (DSP) \cite{Dar2013:opex,Secondini:JLT2013-AIR,Carena:OPEX14,secondini2019JLT,bononi2020fiber}
(and references therein). More recently, the growth of data traffic
driven by artificial intelligence has led to increasing investments
and performance requirements for data center interconnects (DCIs),
where coherent detection and DSP are now being adopted also for short-reach
links. For instance, unrepeatered single-span links are often preferred
for DCIs up to a few hundreds kilometers, provided that more sophisticated
DSP, advanced amplification schemes, and higher launch powers are
used to extend reach without compromising spectral efficiency \cite{agrell2024roadmap}.
As a result, fiber nonlinearity has become a relevant impairment also
in this scenario, and its compensation remains an active area of industrial
interest.

Among the various approaches proposed for modeling and mitigating
fiber nonlinearities, one of particular interest in this context is
the model that treats nonlinear interference (NLI) as a phase noise
process with limited temporal and spectral coherence \cite{Secondini:JLT2013-AIR,secondini2019JLT}.
This representation provides a convenient basis for developing low-complexity
mitigation strategies \cite{civelli2024JLT_CBESSFM},and can further
benefit from subcarrier multiplexing, which provides a simple way
to balance the temporal and spectral coherence of phase noise within
each subcarrier \cite{secondini2019JLT}. Indeed, subcarrier multiplexing
is already widely used in large-bandwidth transceivers because of
its practical advantages \cite{guiomar2017nonlinear,cantono2018networking,welch2023digital}.
This motivates the design of a carrier phase recovery (CPR) scheme
that, besides handling laser phase noise \cite{pfau2009BPS,magarini2012pilot},
also addresses the phase fluctuations induced by fiber nonlinearity
\cite{Neves:24}.

In this paper, we propose a two-stage joint-subcarrier CPR scheme
that operates jointly across all subcarriers of an optical channel
to estimate and compensate the phase noise affecting them. The first
stage mitigates the rapidly varying but deterministic nonlinear phase
distortion arising from intra-channel NLI, while the second stage
uses pilot symbols to estimate the slower stochastic phase noise component
due to inter-channel NLI. We derive the computational complexity of
the proposed algorithm and evaluate its performance through numerical
simulations of an unrepeatered 250\,km fiber link with distributed
Raman amplification.

\section{Joint subcarrier phase recovery scheme}

\begin{figure*}
\begin{centering}
\includegraphics[width=1\textwidth]{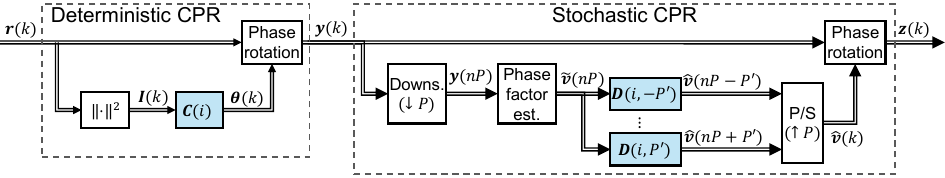}
\par\end{centering}
\caption{\foreignlanguage{american}{\label{fig:schema}Joint-subcarrier CPR: the first stage removes deterministic
phase rotations, while the second stage mitigates stochastic phase
noise. Bold symbols denote vectors collecting the $M$ subcarrier
components. Blue blocks represent MIMO filters.}}
\end{figure*}

The proposed scheme is illustrated in Fig.~\ref{fig:schema} and
consists of two cascaded stages. The first stage, referred to as \emph{deterministic
CPR}, compensates for deterministic nonlinear phase distortions induced
by intra-channel NLI, following the dual-polarization multiband model
proposed in \cite{civelli2024JLT_CBESSFM}. The input vector $\mathbf{r}(k)=[r_{1}(k),\ldots,r_{M}(k)]$
collects the $M'=M/2$ pairs of complex samples of the received signal---$M'$
subcarriers, each with two polarizations---at discrete time $k$,
after matched filtering, symbol-time sampling (at symbol rate $R_{s}$),
and group-velocity dispersion compensation. The subcarrier intensities
$\mathbf{I}(k)=[|r_{1}(k)|^{2}+|r_{2}(k)|^{2},\ldots,|r_{M-1}(k)|^{2}+|r_{M}(k)|^{2}]$
are evaluated and processed by a MIMO filter with $M'\times M'$ matrix
impulse response $\mathbf{C}(i)$, where $i=-N'_{c},\ldots,N'_{c}$
is the time lag, to produce the estimated phase rotations $\bm{\theta}(k)=[\theta_{1}(k),\ldots,\theta_{M'}(k)]$.
Each subcarrier sample is then rotated by the corresponding phase
estimate, which is applied identically to both polarizations.

The second stage, referred to as \emph{stochastic CPR}, mitigates
stochastic phase noise arising from laser phase noise and inter-channel
NLI by exploiting pilot symbols and accounting for correlations over
time as well as across subcarriers and polarizations. The signal vector
$\mathbf{y}(k)$ at the output of the first stage is downsampled by
a factor $P$ and its components are divided by the corresponding
pilot symbols to obtain the preliminary estimates $\tilde{\mathbf{v}}(nP)=[\tilde{v}_{1}(nP),\ldots,\tilde{v}_{M}(nP)]$,
where $\tilde{v}_{m}(nP)=y_{m}(nP)/x_{m}(nP)$. These represent noisy
estimates of the complex phase factors $v_{m}(nP)=\exp(j\phi_{m}(nP))$,
where $\phi_{m}(k)$ is the residual (after the first CPR stage) phase
noise affecting the $m$th component of the received signal at time
$k$. The preliminary estimates are then smoothed and interpolated
by a bank of $P=2P'+1$ (assumed odd for simplicity) parallel $M\times M$
MIMO Wiener filters with matrix impulse responses $\mathbf{D}(i,\ell)$,
where $i=-N_{d},\ldots,N_{d}$ denotes the pilot-time lag and $\ell=-P',\ldots,P'$
the temporal offset between the interpolation time and the nearest
pilot. A subsequent parallel-to-serial (P/S) conversion yields the
final estimates $\hat{\mathbf{v}}(k)=[\hat{v}_{1}(k),\ldots,\hat{v}_{M}(k)]$.
Finally, each component $y_{m}(k)$ is phase-rotated by $\angle\hat{v}_{m}(k)$
to obtain the compensated signal $z_{m}(k)$. For simplicity, pilot
symbols are allocated at the same discrete times ($k=nP$) in all
subcarriers; an optimized allocation strategy is currently under investigation.

\selectlanguage{american}%
The ordering of the two stages is chosen to maximize estimation accuracy.
The deterministic CPR stage compensates for a deterministic phase
contribution that depends only on the signal intensities and is therefore
independent of stochastic phase noise; consequently, its estimation
would not benefit from a prior removal of the latter. In contrast,
the stochastic CPR stage relies on averaging preliminary phase estimates
obtained from pilot symbols and is therefore sensitive to fast phase
fluctuations. If applied before the deterministic stage, these estimates
would be corrupted by the rapid deterministic phase variations induced
by intra-channel NLI, degrading the averaging process. When applied
after after the deterministic stage, these fluctuations are removed,
enabling a more accurate estimation of the residual stochastic phase.
The intuition is confirmed by the numerical results presented in the
next section.

\selectlanguage{english}%
The key components of the proposed two-stage CPR are the MIMO filters
(highlighted in blue), whose coefficients are optimized using training
sequences and a least-squares criterion. The first-stage filter $\mathbf{C}$,
which operates at the symbol rate $R_{s}$, has a long response ($N_{c}=2N_{c}'+1=501$
in this work) and is more efficiently implemented in the frequency
domain via FFT/IFFT pairs and the overlap-and-save technique with
block length $N_{\mathrm{FFT}}\ge N_{c}$. Conversely, the second-stage
filters $\mathbf{D}(i,\ell)$, which operate at the reduced rate $R_{s}/P$,
are much shorter ($N_{d}=2N_{d}'+1=5$ in this work), enabling a straightforward
time-domain implementation. The complexity of the two stages---expressed
in terms of number of real multiplications per complex symbol (RM/2D)---is
\begin{align*}
C_{\mathrm{NLPC}} & =\frac{1}{2}\frac{N_{\text{FFT}}}{N_{\text{FFT}}-2N'_{c}}\left(\log_{2}N_{\text{FFT}}+\frac{3M'+13}{2}\right)\\
C_{\mathrm{PPNC}} & =4M'N_{d}+3
\end{align*}
The latter can be further reduced through suitable approximations
of the matrix impulse responses $\mathbf{D}(i,\ell)$, for example
by pruning small coefficients or exploiting the actual or approximated
symmetries that can be observed in the filter bank. Alternatively,
an FFT-based implementation may also be considered, particularly for
larger $N_{d}$. These additional optimization options are left for
future work.

\section{Performance analysis}

The performance of the proposed scheme is assessed through numerical
simulation of an unrepeatered 250\,km single-mode fiber link (attenuation
0.2\,dB/km, nonlinear coefficient 1.27\,W\textsuperscript{-1}km\textsuperscript{-1},
dispersion 17\,ps/nm/km) with counterpropagating distributed Raman
amplification. The Raman pump power is adjusted in the range 477--663\,mW
to obtain a received signal power of -15\,dBm per channel (sufficient
to ensure no SNR degradation due to the receiver noise) for all considered
signal launch powers.  The WDM signal is composed of 5x180\,GBd
channels spaced by 200\,GHz. Each channel consists of $M'=4$ equally-spaced
subcarriers, each modulated at 45\,GBd by a dual-polarization probabilistic-amplitude-shaping
64-QAM signal employing frequency-domain root-raised-cosine pulses
(roll-off 0.05) and a rate of 4\,bit/2D. At the receiver, after demultiplexing
the central channel, each subcarrier undergoes dispersion compensation,
matched filtering, and symbol-time sampling. Next, either ideal mean
phase removal (MPR) or the proposed CPR scheme is applied, and the
received SNR is estimated (pilot symbols, when used, are excluded
from the SNR evaluation). Simulations are carried out both with and
without laser phase noise, assuming a linewidth of 100\,kHz for both
the transmitter and local-oscillator lasers. The first (deterministic)
CPR stage is implemented with filter length $N_{c}=501$ and block
length $N_{\mathrm{FFT}}=2048$, resulting in a complexity of about
16~RM/2D. The second (stochastic) CPR stage is implemented with
filter length $N_{d}=5$ and pilot periodicity $P=32$, resulting
in a complexity of about 83~RM/2D.

\begin{figure}
\centering{}\includegraphics[width=1\columnwidth]{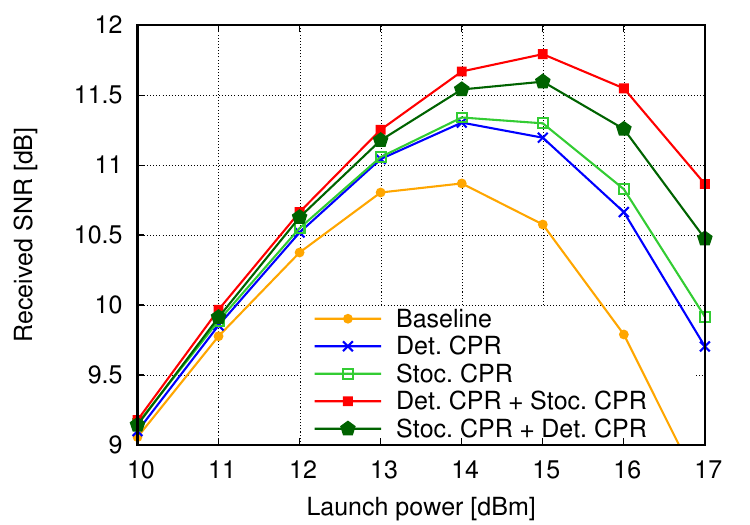}\caption{\foreignlanguage{american}{\label{fig:performance}\foreignlanguage{english}{System performance
without laser phase noise.}}}
\end{figure}

Fig.\,\ref{fig:performance} shows the performance of the proposed
scheme in the absence of laser phase noise, in order to assess its
capability to compensate for fiber nonlinearities. The baseline (yellow
curve) corresponds to simple dispersion compensation followed by ideal
mean phase removal. The remaining curves refer to the proposed CPR
scheme under different configurations: deterministic stage only; stochastic
stage only; and the two stages cascaded either in the order shown
in Fig.\,\ref{fig:schema}, or in the reverse order. 
 The two CPR stages, when used individually, provide similar gains
(0.45\,dB). When cascaded in the order proposed in Fig.~\ref{fig:schema},
they achieve a total gain of 0.9\,dB, confirming that they operate
in a complementary manner by mitigating different contributions of
NLI: the first stage compensates for the deterministic phase distortion
induced by intra-channel NLI, whereas the second stage mitigates the
residual stochastic phase noise arising from inter-channel NLI. Conversely,
when the two stages are cascaded in reverse order, the total gain
is lower (0.7~dB). This confirms the intuition that removing fast
deterministic phase fluctuations before the pilot-aided stage improves
the accuracy of the subsequent estimation and mitigation of the residual
stochastic phase noise.

\begin{figure}
\centering{}\includegraphics[width=1\columnwidth]{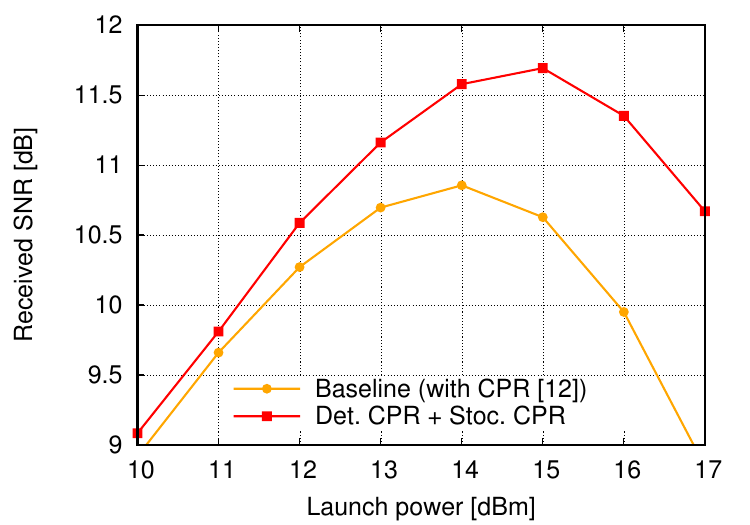}\caption{\foreignlanguage{american}{\label{fig:performance-with_laser_phase_noise}\foreignlanguage{english}{System
performance with laser phase noise.}}}
\end{figure}

The stochastic stage of the proposed CPR is also effective against
laser phase noise. This is illustrated in Fig.~\foreignlanguage{american}{\ref{fig:performance-with_laser_phase_noise}},
which shows the performance of the same system in the presence of
laser phase noise.
 For clarity, we report only the baseline---now including a conventional
pilot-aided CPR based on \cite{magarini2012pilot}, employing a Gaussian
interpolating filter with optimized bandwidth, applied independently
to each subcarrier and averaged across polarizations---and the proposed
scheme in its best configuration, with the two stages cascaded as
in Fig.~\ref{fig:schema}. The inclusion of a conventional CPR in
the baseline is necessary, since under Wiener laser phase noise the
carrier phase follows an unbounded random walk, precluding reliable
operation without CPR; it also reflects the typical configuration
adopted in practical systems. While this conventional CPR effectively
mitigates laser phase noise, it has only a marginal impact on nonlinear
phase distortions, resulting in a performance nearly unchanged with
respect to the case without laser phase noise and without CPR. This
is because the small penalty introduced by laser phase noise is approximately
compensated by a slight gain provided by the conventional CPR in mitigating
nonlinear effects. In contrast, the proposed two-stage CPR simultaneously
mitigates all phase noise contributions, achieving an overall performance
that is only slightly degraded with respect to the previous figure.

\section{Conclusion}

This work presented a low-complexity CPR scheme that compensates for
both laser phase noise and fiber nonlinearity. The proposed scheme
operates jointly across all subcarriers and consists of two cascaded
stages: the first stage compensates for deterministic nonlinear phase
distortions caused by intra-channel NLI (both within and across subcarriers),
while the second stage employs pilot symbols and a bank of Wiener
filters to estimate and mitigate the remaining stochastic phase fluctuations
induced by inter-channel NLI and laser phase noise. In a WDM system
with 4-subcarrier 180\,GBd channels over a 250\,km single-span link
with counterpropagating Raman amplification, the proposed approach
provides a gain of 0.9\,dB with 99\,RM/2D and a pilot rate of 1/32.
The second stage, which accounts for most of the complexity (83\,RM/2D),
can be further improved in terms of both performance (through optimized
pilot allocation across subcarriers) and complexity (by pruning negligible
coefficients and exploiting approximate symmetries in the filter bank).
These refinements are currently under investigation.\pagebreak{}

\printbibliography

\end{document}